


\def\sqr#1#2{{\vcenter{\vbox{\hrule height.#2pt
        \hbox{\vrule width.#2pt height#1pt \kern#1pt
           \vrule width.#2pt}
        \hrule height.#2pt}}}}
\def\square{\mathchoice\sqr34\sqr34\sqr{2.1}3\sqr{1.5}3}
\long\def\UN#1{$\underline{{\vphantom{\hbox{#1}}}\smash{\hbox{#1}}}$}
\def\NL{\hfill\break}
\def\NI{\noindent}
\magnification=\magstep 1
\overfullrule=0pt
\hfuzz=16pt
\voffset=0.1 true in
\vsize=9.2 true in
   \def\NP{\vfil\eject}
   \baselineskip 20pt
   \parskip 6pt
   \hoffset=0.1 true in
   \hsize=6.2 true in
\nopagenumbers
\pageno=1
\footline={\hfil -- {\folio} -- \hfil}

\hphantom{AA}

\hphantom{AA}

\centerline{\UN{{\bf Fast-diffusion mean-field theory for k-body
reactions in one dimension}}}

\vskip 0.4in

\centerline{\bf Vladimir~Privman$^{a,b}$\ \ {\rm
and}\ \ Marcelo~D.~Grynberg$^{a}$}

\vskip 0.2in

\item{$^a$}{\sl{Department of Physics, Theoretical Physics, University
of Oxford, \NL 1 Keble Road, Oxford \ OX1 3NP, UK}}

\item{$^b$}on leave of absence from {\sl{Department of Physics, \NL
Clarkson University, Potsdam, New York \ 13699--5820, USA}}

\vskip 0.4in

\NI {\bf PACS:}$\;$ 05.70.Ln, 68.10.Jy, 82.20.Mj

\vskip 0.4in

\centerline{\bf ABSTRACT}

We derive an improved mean-field approximation for $k$-body
annihilation reactions $kA \to {\rm inert}$, for hard-core diffusing
particles on a line, annihilating in groups of $k$ neighbors with
probability $0 < q \leq 1$. The hopping and annihilation processes are
correlated to mimic chemical reactions. Our new mean-field theory
accounts for hard-core particle properties and has a
larger region of applicability than the standard chemical rate equation
especially for large $k$ values. Criteria for validity of the
mean-field theory and its use in
phenomenological data fits are derived. Numerical tests are reported
for $k=3,4,5,6$.

\NP

Diffusion limited chemical reactions in low dimensions
have attracted much theoretical interest in recent years [1-8].
Indeed, in low dimensions fluctuation effects are more profound;
deviations from the mean-field rate-equation behavior have been
observed in many cases. Most such studies have been focused on two-body
reactions $A+A \to {\rm products}$, and $A+B \to {\rm products}$,
with complications such as particle input and production
(fragmentation), etc.~[8-9].

The $k$-body decay reactions,

$$kA \to {\rm inert} \;\; , \eqno(1) $$

\NI{---} the subject of
our present work, --- have attracted less attention [3]. Indeed, in
actual chemical applications the two-body reactions are essentially the
only relevant processes. However, recently evidence has been
offered that $k$-body reaction kinetics is asymptotically equivalent
to the dynamics of empty sites in certain deposition processes with
diffusional relaxations. Thus, we find it useful to reexamine the
limits of validity of the simplest rate equation corresponding
to (1),

$$ {d c \over dt } = - \Gamma c^k \;\; , \eqno(2) $$

\NI where $\Gamma$ is the phenomenological rate constant, while
$c$ is the particle concentration.

Our study is limited to one-dimensional ($1d$) reactions for
three reasons. Firstly, as
already mentioned, fluctuation effects are most profound in $1d$.
Secondly, numerical simulations of sufficiently high quality are
possible in $1d$, and with modern computer facilities
the $k$ values accessible are $k \leq {\sl O}(10)$. Our
numerical results were obtained with $k=3,4,5,6$. Thirdly, we are going
to argue that the rate equation (2) applies only for times which
increase with $k$ according to

$$ t \gg {\sl O} \left( {\sl e}^k / k^2 \right) \;\; . \eqno(3)$$

\NI This expression applies for large $k$ (and $d=1$);
a more accurate estimate is
given as relation (6) below. Thus, we are going to develop a
modified, improved mean-field theory, following similar ideas advanced
for deposition models [10], with a wider range of applicability:
$t \gg {\sl O}(1)$ for all $k=4,5,\ldots$.
This mean-field theory will be formulated for $1d$ lattice reactions.
Implication of our results for $d \geq 1$ will be discussed in the
summary paragraphs.

Our ``microscopic'' dynamical rules are defined as follows. At a given
time $t$, let the particle density be $\rho (t) = b c(t) $
measured per site of
the $1d$ lattice of spacing $b$. Thus, $\rho$ is dimensionless. We
assume that hard-core particles hop on the lattice and annihilate with
probability $q$, where $0<q \leq 1$, in groups of $k$ neighbors.
Specifically, we assume that each particle attempts to hop at the rate
$H$ per unit time. The hopping direction is selected at random to the
left (rate $H/2$) or to the right (rate $H/2$). The latter case is
illustrated in Figure 1. The active particle marked {\bf a} attempts to
hop to the right. If the neighbor site in the hopping direction is
empty, marked by {\bf e} in Figure 1, then the active particle moves
one lattice spacing. Otherwise it remains in place. However, after each
such hopping attempt, successful or unsuccessful, the active particle
{\bf a} can annihilate with probability $q$ with $k-1$ particles
\UN{\it in the direction of the hopping attempt},
provided of course that all the $k-1$ appropriate consecutive lattice
sites are indeed already occupied.

These dynamical rules introduce correlations between hopping
(diffusion) and reaction. Thus, in fact they are less well described
by mean-field theories than systems in which the microscopic diffusion
and reaction processes are uncorrelated. However, our rules are more
appropriate for mimicing the actual chemical systems in higher
dimensions. Indeed, if we consider diffusion as a result
of Brownian activation due to surrounding medium, then the same
activation should promote the particle cluster to go over
the reaction energy barrier.

The mean-field approximation is introduced by assuming that the effect
of diffusion is to eliminate all correlations in particle positions.
For one-dimensional hard-core particles (without
any reactions), such diffusional
``decorrelation'' occurs for infinite times [4].
For reacting particles the
approximation is therefore exact for infinite rate of diffusion
as compared to reaction, i.e., in the limit $q \to 0$. In $d=1$
the fast-diffusion approximation takes on a
particularly simple form [10].
Indeed, the particle positions being uncorrelated means that the
(normalized)
probability to a find gap of $m=0,1,2, \ldots$ lattice spacings between
two consecutive particles is

$$ {\rm Prob} (m) = \rho ( 1- \rho )^m \;\; , \eqno(4) $$

\NI independent of the positions of the nearby particles.

The annihilation event shown in the upper panel of Figure 1 will occur
with probability $\rho (1- \rho ) \times \rho^{k-2}$, while the event
shown in the lower panel will occur with probability
$\rho \times \rho^{k-2}$, where we assume that the gap distribution is
given by (4). The factors $\rho (1-\rho )$ and $\rho$ are the
probabilities to find gaps of size 1 and 0, respectively, while
the factors $\rho^{k-2}$ ensure that the next $k-2$ gaps
are all 0.
Thus the rate of annihilation events \UN{\it per site\/} will
be the sum of the above probabilities times the fraction of occupied
sites, $\rho$, and the overall rate of annihilation per site, $qH$.
The decrease in the particle
density is therefore described by the relation

$$ {d \rho \over dt } = - q k H (2 - \rho ) \rho ^k \;\; , \eqno(5) $$

\NI where the factor $k$ was introduced because in each annihilation
$k$ particles are removed.

The derivation reported above applies for an infinite system.
With some more combinatorics, one can also derive the mean-field
equation appropriate for a finite system of $N$ lattice
sites. This development has not been attempted here.

Having derived the fast-diffusion mean-field approximation, relation
(5), for the variation of the particle
density, we proceed as follows. We will
derive the limits of validity of the mean-field theory, followed by
a discussion of how it should be used in actual data fitting. We then
report numerical tests for $k=3,4,5,6$. Two comments are,
however, in place at this stage.

Firstly,
the precise form of the mean-field approximation depends on the
microscopic dynamical rules. Indeed, the
factors combining to produce the right-hand side of (5) were obtained
by considering the specific annihilation events presented in Figure 1.
Any change in the rules may modify the form of (5), although the
general proportionality to $\rho^k$ for small $\rho$ will apply because
$k$ particles participate in each annihilation. However,
our discussion below, of various aspects of the use and limits of
validity of the mean-field approximation, applies quite generally.

Secondly, for $\rho << 1$
the form (5) reduces to the rate equation (2) with the rate
constant $\Gamma_{\rm bare} = 2qkb^{k-1}H$, where the subscript will be
explained later. However, one can show that the precise condition for
attaining this regime is

$$ t \gg {{\sl e}^{k-1} \over 2qHk(k-1) } \;\; . \eqno(6) $$

\NI Thus, the simple rate equation (2) fails for times increasing
rapidly with $k$, even within the limits of validity of the
fast-diffusion approximation (which will be described shortly). It
is important to point out, however, that modification of the
simplest form (2) needed to extend the time range down to
$qHt \gg {\sl O} (1) $,
depends on the precise microscopic dynamical rules.

We now derive the expression for the limits of validity of
mean-field theory. Our approach differs from that adopted in [3] and
follows instead the ideas developed for deposition processes [10].
The result is however qualitatively the same.
Mean-field relations apply in those regimes
where the \UN{\it local\/} density fluctuations do not dominate the
evolution of the reaction, but instead it is controlled by the global
variation in various average properties. As the first step, let us
consider in what regimes the local density fluctuations are negligible
in their effect on the reaction kinetics.

The average distance between two consecutive particles is
$b (1-\rho )/\rho$, so that $k$ particles explore diffusively the
distance $\ell = kb(1-\rho )/\rho $. Annihilation rate in this
length is given (within the mean-field approximation) by

$$ \omega = qH(2-\rho ) \rho^k \times (\ell /b )
= k qH (1- \rho ) (2- \rho ) \rho^{k-1} \;\; , \eqno(7) $$

\NI where the factors multiplying $(\ell /b)$ in the intermediate step
represent the annihilation rate (per unit time) per site.

Each annihilation event perturbs the density distribution over the
distance of order $\ell$,
which ${\sl O} (k)$ particles would explore had there been no
annihilation at all. Thus, we expect that for the fast-diffusion
approximation to apply the diffusion must smooth out density
fluctuations in $\ell$ much faster than the annihilation events
occur. Now the time scale for diffusional relaxation, $\tau$,
is given by

$$ D \tau = \ell^2 =
\left[kb(1-\rho )/\rho \right]^2  \;\; , \eqno(8) $$

\NI where $D \equiv b^2 H /4 $ is the diffusion constant of the
particles in the dilute limit.

Mean-field theories can be used provided

$$ \omega \tau \ll 1 \;\; , \eqno(9) $$

\NI which, after collecting all the definitions and relations above,
yields

$$ {1 \over q} \gg [2k(1- \rho )]^3 \rho^{k-3} \;\; ,
\eqno(10) $$

\NI where we replaced $\left( 1- {\rho \over 2} \right)$
by 1 on the right-hand side; this factor is of order 1 in all regimes.

The right-hand side of (10) is plotted in Figure 2 for $k=2,3,4,5$.
Asymptotically, for small densities (i.e., for large times) the
mean-field approximation will always fail for $k=2$. For $k=4,5,\ldots$
the mean-field theory provides the correct asymptotic description of the
large-time reaction kinetics. The case $k=3$ is borderline.

The precise nature of the approximation by the mean-field relations
must be discussed in greater detail, however. Indeed, strictly speaking
the mean-field results are exact only when correlations are absent
which in our case corresponds to the limit $q\to 0$, i.e., very slow
reaction. For any finite $q$, there are always correlations. For our
choice of the dynamics some correlations are in fact easily visualized.
For instance, for $q=1$, the configuration shown in the lower panel
of Figure 1 will be never generated dynamically because \UN{\it all\/}
hopping events forming sequences of $k$ or more particles are followed
by reaction events. The only
source of large occupied regions will be the unreacted sequences of
particles left over from the initial state at time $t=0$. The gap
size distribution (4) cannot be exact for systems with
such correlations (which are also partially present for $q<1$).

More generally, depletion and distortion of two- and
multiple-particle correlation functions away from their mean-field
values occurs
due to correlations in the dynamics. As a result, the mean-field
expressions can be used only with \UN{\it effective}, renormalized
(also termed ``hydrodynamic'')
rate parameters, rather than with \UN{\it bare\/} (also known as
coarse-grained) rate parameter
values such as $H$, even in the regimes where they provide the correct
functional form of the asymptotic dynamics.

Specifically, we can define the effective rate function

$$ R_{\rm eff} (t) \equiv {1 \over qk (\rho -2) \rho^k }
{d \rho \over dt } \;\; , \eqno(11) $$

\NI suggested by the form of the mean-field relation (5). We expect
this function to assume constant values, say $R$, in all the
asymptotically mean-field regimes,
where $R \neq H$. Since fast reaction tends to decrease the probability
of small interparticle gaps, we expect the renormalized rate
parameter $R$ to satisfy $R \leq H$, with equality for $q \to 0$.

In our derivation of the criterion (10) for the applicability of the
mean-field theory, we used the length scale $\ell =
kb(1-\rho)/\rho$ which
becomes small for $\rho \to 1$. A conclusion suggested by Figure 2,
that the mean-field relations apply for $\rho \simeq 1$, is valid only
in a state well equilibrated diffusively. Thus, for initial densities
of order 1, our criterion can be used only for \UN{\it
random initial particle
distribution}, or for nonrandom distributions which only involve
correlations over lengths smaller than $kb[1-\rho(0)]/\rho(0)$.

Thus, there are two possibilities for the mean-field expression to
apply for short times, provided that the initial distribution is
sufficiently random. If the $\rho(0)$ and $q$ values are such that
the inequality (10) is well satisfied to the right of the ``hump'' of
the curves $k \geq 4$ in Figure 2 (or to the right of the full
curves for $k=2,3$), then the effective rate parameter
$ R_{\rm eff} (t) $ will be initially more or less constant. However,
this constant value need not be the same as the one attained for large
times, to the left of the ``hump,'' for $k=4,5,\ldots$,
and in fact we expect it to be larger than the $t=\infty$ value.
If, however, the initial density is sufficiently small,
or $1/q$ sufficiently large,
then the whole density variation may fall in the low-density
regime (for $k\geq 4$), in which case the mean-field relation with the
effective $R$ value replacing $H$, will apply for all times.

The rate function $ R_{\rm eff} (t) $ should therefore decrease
monotonically with time. If the initial density is close to $1$, and
the distribution random, then the rate function will
have a distinct plateau for short times, with values approaching
$H$ for $\rho (0) \to 1$. For large times the rate function will
decrease to the asymptotic mean-field renormalized rate parameter value
$R$, provided $k=4,5,\ldots$. If the initial density is sufficiently
small, or $1/q$ sufficiently large, then this large time behavior
will in fact set in for short times as well.

For $k=2$ the mean-field theory breaks down asymptotically for
sufficiently large times, see
Figure~2. The rate function then decreases to 0
according to the power law $\sim 1/\sqrt{t}$, as follows from results
available in the literature [2,5-7]. Generally for reaction kinetics
with identical particles, the nature of deviation from the
mean-field behavior for large times can be related to the problem of
repeated
meetings of $k$ random walkers; see, e.g., [7]. However, presently
all such mathematical results are limited to two-body reactions.
For short times and fast diffusion,
the $k=2$ system can have certain mean-field like
properties which, however, fade away as time increases. This behavior
has been observed numerically for reactions [11] and for related
deposition models [12].

For $k=3$, --- the borderline case,
--- it is likely that the rate function vanishes logarithmically for
large times, $\sim 1/\log t$, though we are not aware of any
published results to substantiate this expectation.
However for any data set for
$k=3$, taken over several decades of $t$, the
mean-field relations can be phenomenologically used with the effective
rate parameter $R>0$. The marginal logarithmic vanishing of the rate
function is difficult to observe numerically.

Our Monte Carlo data were obtained for $k=3,4,5,6$, on lattices of
sizes 2000 with periodic boundary conditions. Each data
set was averaged over 50 independent runs, all with random
initial conditions. A few runs, order 10, were also made for
$k=10$. We replace $H$ by $R$ in the mean-field
approximation (5) and all relations that derive from it.
The Monte Carlo
time steps were selected to have the microscopic hopping attempt rate
$H=1$. Thus we expect the effective rate values to satisfy $R<1$.
Our typical numerical results are illustrated in Figures 3-6.

Integration of the mean-field relation yields the function
$t(\rho )$ as follows,

$$ q 2^k k R \left[ t(\rho ) - t_0 \right] = I(\rho ) - I(\rho_0)
+ \ln {2- \rho \over 2 - \rho_0 } \;\; , \eqno(12) $$

\NI where

$$ I(\rho ) \equiv -\ln \rho + \sum_{j=1}^{k-1} {2^j \over j \rho^j}
\;\; , \eqno(13) $$

\NI provided that mean-field theory applies for times $t \geq
t_0$, with the corresponding density $\rho_0 = \rho (t_0)$. As
exemplified by Figures~3-6, our data eventually reached the large-time
behavior not sensitive to the initial density. This observation
suggests that
the large-time value $R$ only depends on $q$, but not on $\rho(0)$.
Thus, we used this large time data to fit the $R$ value from the
relation (12) in which the terms which are of order 1 for
small densities were discarded,

$$ R t \simeq I(\rho )/q 2^k k \;\; . \eqno(14) $$

\NI These curves are shown by dashed lines in Figures 4 and 6, while in
Figures 3 and 5 they were too close to other lines (solid lines marked
$0.8$) to be shown. On a double-logarithmic plot, variation of the
trial $R$ value corresponds to translation of the dashed
line along the $\, \log t \, $ axis. Thus, we only obtain the estimates
of $\, \log R$. The accuracy of the resulting $R$ values is at best
semiquantitative.

The values of the large-time asymptotic rate constant, $R$,
are summarized in Table 1. The overall trend is as expected from our
heuristic discussion of the validity of the mean-field approximation.
The effective rates $R/H$ approach $1$ for small $q$, while for
$q \simeq 1$ the mean-field relation applies with substantially
renormalized values $R<H$. Numerical data for several decades in $t$
cannot be used to detect logarithmic terms for $k=3$.
However, the $k=3$ estimates of $R$
in Table 1 are markedly lower than their
$k>3$ counterparts with the same $q$.

The dashed lines in Figures 3 and 5, defined by (14), deviate in two
ways from similar relations predicted by the simpler rate equation
(2) with $\Gamma = 2qkb^{k-1} R$ (which are not shown in the figures).
Firstly, the curves differ
for short times. A more interesting observation is that
for larger times, i.e., for smaller $\rho$ values,
the dashed lines in Figures 3 and 5 look nearly straight. However,
their slope is somewhat steeper than the prediction of the
asymptotic rate equation (2): slope $-1/(k-1)$. Indeed, this
deviation is quite small for $k\leq 6$. It becomes more profound
as $k$ increases, as was found in our preliminary, limited-statistics
runs for $k=10$ as well as in Monte Carlo simulations [13] of
related deposition models up to $k=10$.
Of course, asymptotically the slope slowly approaches the rate-equation
value, for times defined by (6).

Statistical noise in our data precluded direct estimation of the
rate function (11) because evaluation of the time derivative turns
out to be particularly sensitive to statistical fluctuations. However,
we used the $R$ values estimated for large times (Table 1) and the
initial values $\rho(0)$ to draw mean-field curves (12) corresponding
to different initial densities. As expected, the short-time behavior
of the data is fitted well only for small $\rho(0)$; the quality of the
fit improves as $q\to 0$. These properties are illustrated in Figures
3-6 (solid curves). They were shared by all our data sets listed
in Table 1.
As discussed earlier, the mean-field theory either fails for short
times or applies with the effective rate constant larger than $R$,
unless the initial density and $q$ are both sufficiently small.
Only in the latter case the fixed-$R$
mean-field approximation extends down to $t=0$.

In summary, we analyzed the applicability of a mean-field
approximate equation accounting for the hard-core particle dynamics,
to chemical reactions in $1d$. Some of our conclusions are generally
valid for $d \geq 1$; these include the fact that difficulties with
the simplest rate equations for large $k$ are not inherent to
mean-field approximations and can be repaired by accounting for the
hard-core interactions, although our explicit results were limited to
the one-dimensional case.

Another well known
general feature illustrated by our $1d$ studies, is that
mean-field theories break down in those
cases when local fluctuations dominate the dynamics of the reaction.
Classification of borderline $d$-values at and below which the
mean-field theory breaks down for multiparticle-input reactions
$k_1 A_1 + k_2 A_2 + k_3 A_3 + \ldots \to {\rm inert}$, etc., by
scaling agruments, can be found in [3,14].

However, even in the regimes where the local
fluctuations are irrelevant
asymptotically, the $1d$ model studies emphasize the fact
that mean-field theories can only be applied
as ``effective'' asymptotic approximations, with
renormalized, ``hydrodynamic'' rate parameters. Our study further
suggests that with careful choice of a mean-field equation,
one-parameter data fits work quite adequately for large times and
in some cases describe the behavior down to $t=0$.

The authors wish to thank M.~Barma for
helpful comments and suggestions.
This research was partially supported by the
Science and Engineering Research Council (UK)
under grant number GR/G02741.
One of the authors (V.P.) also wishes to acknowledge
the award of a Guest Research Fellowship at Oxford from the Royal Society.

\NP

\centerline{\bf REFERENCES}

{\frenchspacing

\item{1.} M. Bramson and D. Griffeath,
Ann. Prob. {\bf 8}, 183 (1980).

\item{2.} D.C. Torney and H.M. McConnell,
J. Phys. Chem. {\bf 87}, 1941 (1983).

\item{3.} K. Kang, P. Meakin, J.H. Oh and S. Redner,
J. Phys. A{\bf 17}, L665 (1984).

\item{4.} T. Liggett, {\sl Interacting Particle Systems\/}
(Springer-Verlag, New York, 1985).

\item{5.} Z. Racz, Phys. Rev. Lett. {\bf 55}, 1707 (1985).

\item{6.} A.A. Lushnikov, Phys. Lett. A{\bf 120}, 135 (1987).

\item{7.} M. Bramson and J.L. Lebowitz, Phys. Rev. Lett. {\bf 61},
2397 (1988).

\item{8.} Review: V. Kuzovkov and E. Kotomin, Rep. Prog. Phys.
{\bf 51}, 1479 (1988).

\item{9.} D. ben--Avraham, M.A. Burschka and C.R. Doering,
J. Stat. Phys. {\bf 60}, 695 (1990).

\item{10.} V. Privman and M. Barma, Oxford preprint
{\sl OUTP--92--24S\/} (1992).

\item{11.} L. Braunstein, H.O. Martin, M.D. Grynberg
and H.E. Roman, J. Phys. A{\bf 25}, L255 (1992).

\item{12.} V. Privman and P. Nielaba, Europhys. Lett.
{\bf 18}, 673 (1992).

\item{13.} P. Nielaba and V. Privman, Modern Phys. Lett. B (1992),
in print.

\item{14.} S. Cornell, M. Droz and B. Chopard, Physica A (1992), in
print.

}
\NP

\centerline{\bf TABLE}

\hphantom{AA}

\NI\hang {\bf Table~1.}~$\;$Large time estimates of the
phenomenological mean-field rate constant $R \leq H$,
based on Monte Carlo data up to $t=10^6H^{-1}$.
Due to statistical noise in the data, the values $R/H$ shown
are semiquantitative; no reliable error limits can be offered.
(The $q=0.001$ Monte Carlo run was only for $k=3$. Limited-statistics
results for $k=10$, up to times $tH=10^7$, were also obtained,
but no reliable $R$ estimates can be offered.)

\vskip 0.20 in

$$\vbox{
\settabs\+ &AAAAAAAAA&AAAAAAAA&AAAAAAAA&AAAAAAAA&AAAAAAAA\cr

\+ & & $k=3$ & $k=4$ & $k=5$ & $k=6$ \cr

\+ & $q=1$ & $0.05$ & $0.13$ & $0.18$ & $0.2$ \cr

\+ & $q=0.1$ & $0.3$ & $0.5$ & $0.6$ & $0.6$ \cr

\+ & $q=0.01$ & $0.7$ & $0.9$ & $0.9$ & $0.9$ \cr

\+ & $q=0.001$ & $0.96$ &\NI{------} &\NI{------} &\NI{------} \cr

}$$

\NP

\centerline{\bf FIGURE CAPTIONS}

\NI\hang {\bf Fig.~1.}~$\;$The active particle {\bf a} attempts to hop
to the right. The upper panel shows a configuration in which the
attempt is successful: the particle will move to the empty site {\bf
e}. The lower panel shows a blocked
configurations. In both cases, the $k-1$ lattice sites in the direction
of the hopping attempt are occupied ($k=5$ here). Thus, all $k$
particles shown will annihilate with probability $q$.

\NI\hang {\bf Fig.~2.}~$\;$The function defined by the right-hand side
of the inequality (10), shown for $k=2,3,4,5$.

\NI\hang {\bf Fig.~3.}~$\;$Numerical data for $k=6$ and $q=1$, with the
initial densities $\rho(0)=0.8$ ($\circ$), $0.5$ ($\triangle$),
0.2 ($\square$). The solid lines, labeled by the $\rho (0)$
values, represent the mean-field relation
(12) with the large-time $R$ value, forced to obey the initial
condition $\rho = \rho(0)$ at $t=0$.

\NI\hang {\bf Fig.~4.}~$\;$Numerical data for $k=6$ and
$q=0.01$, with the
initial densities $\rho(0)=0.8$ ($\circ$), $0.5$ ($\triangle$),
0.2 ($\square$). The solid lines, labeled by the $\rho (0)$
values, represent the mean-field relation
(12) with the large-time $R$ value, forced to obey the initial
condition $\rho = \rho(0)$ at $t=0$.
The dashed line indicates the large-time asymptotic expression (14)
used to fit the $R$ value.

\NI\hang {\bf Fig.~5.}~$\;$Numerical data for $k=3$ and
$q=0.1$, with the
initial densities $\rho(0)=0.8$ ($\circ$), $0.5$ ($\triangle$),
0.2 ($\square$). The solid lines, labeled by the $\rho (0)$
values, represent the mean-field relation
(12) with the large-time $R$ value, forced to obey the initial
condition $\rho = \rho(0)$ at $t=0$.

\NI\hang {\bf Fig.~6.}~$\;$Numerical data for $k=3$ and
$q=0.001$, with the
initial densities $\rho(0)=0.8$ ($\circ$), $0.5$ ($\triangle$),
0.2 ($\square$). The solid lines, labeled by the $\rho (0)$
values, represent the mean-field relation
(12) with the large-time $R$ value, forced to obey the initial
condition $\rho = \rho(0)$ at $t=0$.
The dashed line indicates the large-time asymptotic expression (14)
used to fit the $R$ value.

\hphantom{A}

\NI {\bf To get the preprint with the figures write to \NL
PRIVMAN@CRAFT.CAMP.CLARKSON.EDU}

\bye